\documentclass[12pt,preprint]{aastex}

\usepackage{natbib}
\shorttitle{Detection of Star Formation}
\shortauthors{Megeath et al.}

\bibpunct{(}{)}{,}{a}{}{;}
\bibpunct{(}{)}{,}{a}{}{;}

\begin{document}

\title{Detection of Star Formation in the Unusually Cold Giant
  Molecular Cloud G216\footnote{Based on observations made with ESO
    telescopes at the La Silla Observatory}}


\author{S. T. Megeath\altaffilmark{1}, E. Allgaier\altaffilmark{1}, E. Young\altaffilmark{2}, T. Allen\altaffilmark{1,3}, J. L. Pipher\altaffilmark{3,4} \& T. L. Wilson\altaffilmark{5}}

\altaffiltext{1}{Ritter Observatory, MS-113, University of Toledo, 2801 W. Bancroft St,
Toledo, OH, 43560 (megeath@physics.utoledo.edu)}

\altaffiltext{2}{Steward Observatory, University of Arizona, 933 North Cherry Avenue, Tucson, AZ, 85721}

\altaffiltext{3}{Department of Physics and Astronomy, University of Rochester, Rochester, NY 14627}

\altaffiltext{4}{Visiting Astronomer, Kitt Peak National Observatory, National Optical Astronomy Observatory, which is operated by the Association of Universities for Research in Astronomy (AURA), Inc., under cooperative agreement with the National Science Foundation.}

\altaffiltext{5}{European Southern Observatory, Karl-Schwarzschild-Strasse 2, 85748 Garching, Germany}



\begin{abstract}

The giant molecular cloud G216-2.5, also known as Maddalena's cloud or
the Maddalena-Thaddeus cloud, is distinguished by an unusual
combination of high gas mass ($1-6 \times 10^5$~M$_{\odot}$), low
kinetic temperatures (10~K), and the lack of bright far infrared
emission.  Although star formation has been detected in neighboring
satellite clouds, little evidence for star formation has been found in
the main body of this cloud.  Using a combination of mid-infrared
observations with the IRAC and MIPS instruments onboard the {\it
  Spitzer} space telescope, and near-IR images taken with the
Flamingos camera on the KPNO 2.1-meter, we identify a population of 41
young stars with disks and 33 protostars in the center of the cloud.
Most of the young stellar objects are coincident with a filamentary
structure of dense gas detected in CS ($2 \rightarrow 1$).  These
observations show that the main body of G216 is actively forming
stars, although at a low stellar density comparable to that found in
the Taurus cloud.

\end{abstract}


\section{Introduction}

The weak CO emission from the cold giant molecular cloud G216-2.5 was
first discovered by \citet{1985ApJ...294..231M} in a survey of the
Orion-Monocerous region.  The cloud is unusual because it has the
typical size \citep[$250 \times 100$~pc,][]{1985ApJ...294..231M} and
mass \citep[$1-6 \times 10^5$~M$_{\odot}$,][]{1994ApJ...432..167L} of
a giant molecular cloud (hereafter: GMC), yet it has a kinetic
temperature of only 10~K \citep{1998ApJ...494..657W}.  The low
temperature is unusual for GMCs and is more typical of dense cores in
the Taurus dark cloud \citep{1999ApJS..125..161J}.  Perhaps the most
unusual characteristic of the cloud is the lack of star formation.
\citet{1985ApJ...294..231M} found no clear evidence for star formation
within the G216-2.5 GMC (hereafter: G216); a result supported by
subsequent IRAS maps showing a distinct absence of emission from dust
heated by internal young stars
\citep{1996ApJ...472..275L}. Considering the current distance estimate
of 2.2 kpc \citep{1991ApJ...379..639L}, the absence of bright far-IR
emission only rules out the presence of young high mass stars.  Young
low to intermediate mass stars could have escaped the detection by
IRAS.  Nevertheless, the evident lack of massive star formation is in
itself unusual since GMCs almost ubiquitously contain young massive
stars \citep{1997ApJ...476..166W}.

\begin{figure}
\epsscale{1.2}
\plotone{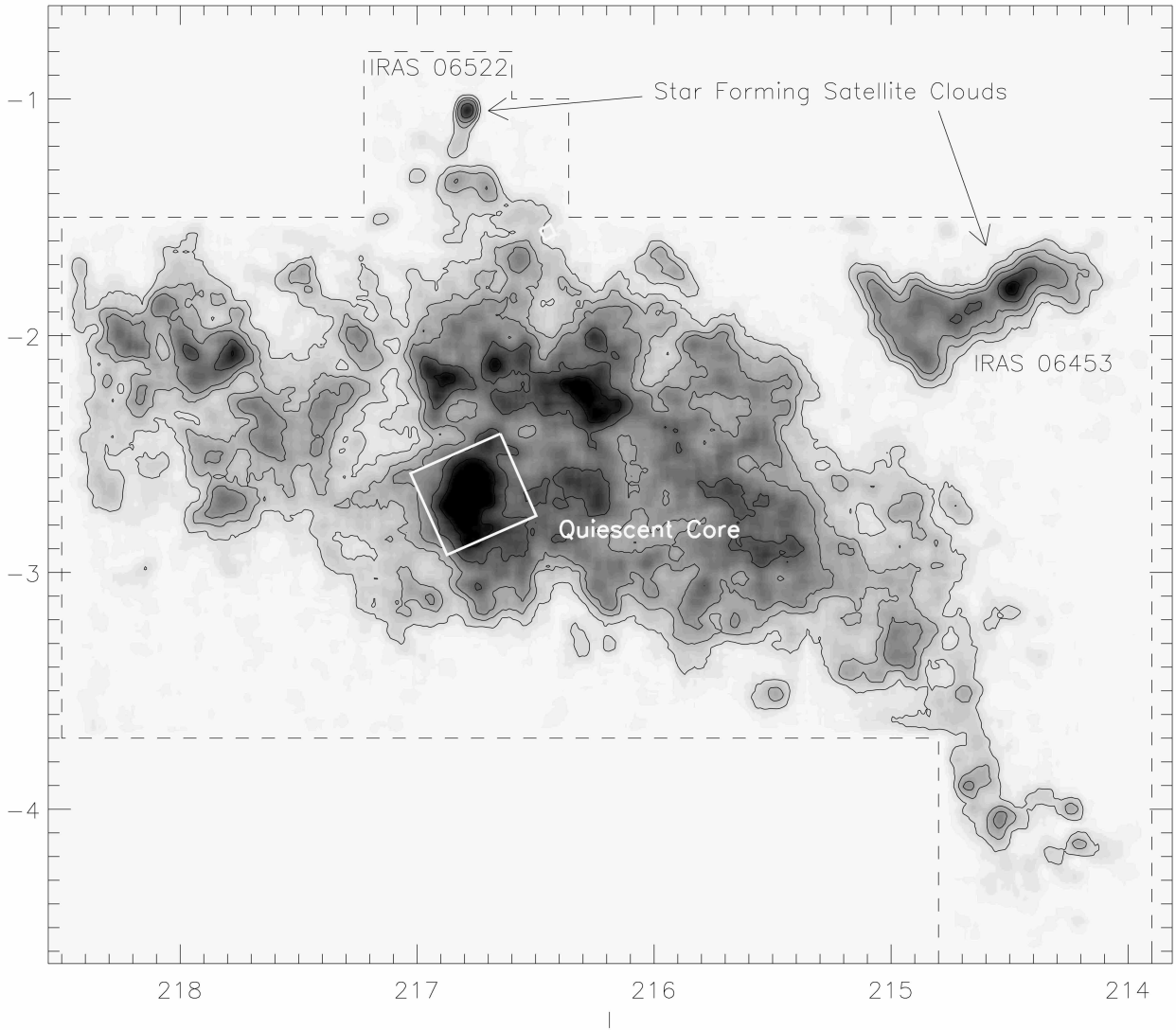}
\caption{CO (1-0) map of G216-2.5 with the location of the Spitzer map
marked in the white box; this map was made from the data of
\citep{1994ApJ...432..167L}.  Areas of star formation previously
identified by \citet{1996ApJ...472..275L} are indicated.}
\label{fig:cloudmap}
\end{figure}

Because of the lack of evident star formation, G216 has been considered
the best example of a quiescent GMC and potentially, a rare example of
a GMC before the onset of star formation \citep{1985ApJ...294..231M}.
In the 1990s, near-IR imaging by \citet{1996ApJ...472..275L} revealed
two small groups of young stars in satellite clouds on the northern
edge of G216 (Fig.~\ref{fig:cloudmap}).  One of these regions is
associated with a radio source/HII region which was first noted by
\citet{1985ApJ...294..231M}, but Maddalena et al. could not reliably
associate the radio source with the satellite cloud.  Near-IR
observations of the main body of the cloud failed to detect any
stellar groups similar to those in the satellite clouds.

\citet{1994ApJ...432..167L} argued that the main body of the cloud was
the site of past star formation.  Their evidence was the the presence
of shell-like structures in the cloud, the existence of a large scale
velocity gradient across the cloud, and a virial mass five times the
mass estimated from the luminosity of the $^{13}$CO emission; all of
these features could be the result of previous episodes of star
formation which stirred up motions in the cloud.  In support of this
idea, HI mapping by \citet{1996ApJ...464..247W} showed that G216 is
linked by a 50~pc photodissociation region to two neighboring
molecular clouds containing young OB stars. In total, these
observations have established that G216 is part of a larger complex;
however, it is not clear whether G216 is a relic of past star
formation or a quiescent component of a larger cloud complex
containing star forming clouds.  To date, the relic scenario has one
major drawback: the lack of a detectable population of stars in the
cloud.  Interestingly, the only sites of previously known star
formation directly associated with G216 are found in satellite clouds
near the surface of G216 that faces the Sh287 star forming region
(Fig.~\ref{fig:cloudmap}).  This geometry suggests that star formation
may have been initiated by interactions with Sh287.

Despite the controversy over the nature of G216, it remains an
unusually cold and quiescent cloud compared to other GMCs with similar
masses.  For this reason, detailed studies of this cloud may offer new
insights into the condition necessary for star formation and the
factors that drive the rate and efficiency of star formation.  We have
recently completed a {\it Spitzer} space telescope survey of the core
of G216.  This core region shows an extended region of relatively high
column density gas \citep{1991ApJ...379..639L,2006ApJ...643..956H},
but previous to this study showed little evidence for ongoing star
formation.  Using {\it Spitzer} and data from the KPNO 2.1 meter and
SEST telescopes, we show that the central core contains a complex of
dense cores undergoing star formation. A total of 33 protostars is
found, indicating that the region is currently undergoing vigorous
star formation. In addition 41 young pre-main sequence stars with
disks are identified.  We compare the rate of star formation in G216
to that in nearby molecular clouds in Orion and Taurus.

\section{Observations \& Data Reduction}

G216 was observed with the IRAC and MIPS instruments onboard the {\it
  Spitzer} space telescope for the GO program PID 3178
\citep{2004ApJS..154...10F,2004ApJS..154...25R}.  The field was mapped
with the IRAC instrument in two separate observations on March 25,
2005. During each observation, images were taken at series of pointing
positions defined by a $5 \times 7$ rectangular grid with a spacing
between adjancent grid points of $260''$.  This resulted in four
overlapping but offset maps in the separate IRAC bands providing
complete four band coverage over a $22' \times 31'$ region.  Three
dithered 30 second frames were taken at every grid position. The high
dynamic range mode was selected; each 30 second frame was accompanied
by a 1.2 second frame.  The total integration time for the two
observations and three dithers was 160.8 seconds. The MIPS data were
obtained in four separate observations: two on October 10, 2004, one
on April 01, 2005 and one on April 08, 2005.  In total, four scan maps
centered on the same position were obtained; these were executed with
the medium scan rate.  Since the direction of the scanning depended on
the date observed, the four maps have three different scan orientations.
This improves the redundancy of the data and minimizes artifacts.  The
scan maps are $25' \times 45'$ in extent with $148''$ spacings between the
scan legs.

Photometric magnitudes in the four IRAC bands were extracted with
PhotVis version 1.10, an IDL GUI-based photometry visualization tool
\citep{2004ApJS..154..374G}.  The aperture size was 2 pixels ($2.4''$)
and the sky annulus extended from 2 ($2.4''$) to 6 ($7.2''$) pixels.
To convert the signal (measured in DN~s$^{-1}$) to magnitudes, we used
zero points of 19.6642, 18.9276, 16.8468 and 17.3909 in the 3.6, 4.5,
5.8 and 8~$\mu$m bands respectively \citep[these are derived from the
  calibration of ][]{2005PASP..117..978R}.  The signal was multiplied
by an aperture correction to take into the small aperture size; the
adopted values were 1.213, 1.234, 1.379, and 1.584 for 3.6, 4.5, 5.8
and 8~$\mu$m bands, respectively. The magnitudes from the 1.2~s frames
were used if the long frame magnitudes were less than 11, 10, 7, and 8
at 3.6, 4.5, 5.8 and 8~$\mu$m, respectively; at this point, a
comparison of the 30 and 1.2~s frame photometry showed that the long
frames were becoming non-linear.  A total of 11,195 sources were
detected in at least the 3.6 and 4.5~$\mu$m bands and 2508 sources
were detected in all four IRAC bands.

The MIPS data were processed using the MIPS instrument team's Data
Analysis Tool \citep{2005PASP..117..503G}, which calibrates the data,
corrects distortions, and rejects cosmic rays during the co-adding and
mosaicking of individual frames.  The MIPS 24~$\mu$m mosaic created
from the four scan maps has a pixel size of $1.245''$ and
units of DN~s$^{-1}$~pix$^{-1}$.  MIPS 24~$\mu$m photometric
magnitudes were extracted using the IDL implementation of DAOPHOT in
the IDL Astronomy User's Library \citep{1993ASPC...52..246L}.  Sources
were first identified with PhotVis.  An initial determination of their
magnitudes was then obtained by aperture photometry with PhotVis; the
aperture was set to 5 pixels ($6.225''$) and the sky annulus extended
from 12 ($14.94''$) to 15 ($18.675''$) pixels.  The zero point
magnitude for the mosaicked 24~$\mu$m data was 16.05. This assumed a
mosaic pixel that was 1/2 the diameter of an instrument pixel, a
calibration factor of $6.4 \times 10^{-6}$~Jy per DN~s$^{-1}$~per
inst.~pix, a zero magnitude flux of 7.17~Jy, and a correction from 12
pixels to infinity of 1.146 \citep{2007PASP..119..994E}.  Since we
measured our photometry with a 5 pixel aperture, a correction from a 5
pixel to 12 pixel aperture was determined from the brighter stars in
our sample; the correction of -0.428~mag was included in the zero
point magnitude.  A PSF with a 15 pixel radius was then generated
using 10 bright, isolated stars scattered around the 24 micron image;
these were selected to be relatively free of contamination from nearby
sources and nebulosity.  The aperture photometry of these 10 stars set
the calibration of the PSF fitting photometry.  The PSF was fit to all
the point sources identified in the image using a 2 pixel fitting
radius (i.e. only pixels in an inner 2 pixel radius were used in the
fit).  The PSF fit to each source was then subtracted and the above
procedure was repeated to identify faint point sources hidden in the
wings of brighter sources.  In the second iteration, 13 additional
sources were found.  A total of 253 sources had detection in all four
IRAC bands and the MIPS 24~$\mu$m band.

The 70~$\mu$m mosaic created from the four scan maps has a pixel size
of $9.9''$ and units of DN~s$^{-1}$.  We performed point source
identification and photometry on the MIPS 70~$\mu$m mosaic with
Photvis. An aperture size of $16''$ and a sky annulus extending from
$18''$ to $39''$ were used.  The adopted zero magnitude was of
$-1.57$; this was based on an aperture correction of 2.07, a
calibration factor of 1.6~Jy~(DN/s/pix)$^{-1}$, and a zero flux of
0.778~Jy \citep{2007PASP..119.1019G}.

Deep near-IR data were obtained with the Flamingos camera on the KPNO
2.1-m telescope on November 2, 2004.  Flamingos data reduction was
performed using custom IDL routines, including modules for
linearization, flat-field creation and application, background frame
creation and subtraction, distortion measurement and correction, and
mosaicking.  Point source detection and synthetic aperture photometry
of all point sources were carried out using PhotVis. Aperture
photometry was performed using radii of 2''.73, 3''.64, and 6''.06 for
the aperture, inner, and outer sky limits, respectively.  The
photometry was calibrated by measuring the magnitude offset between
the instrumental magnitudes derived from the Flamingos data and the
apparent magnitudes from the 2MASS point source catalog.  For the
$K$-band data, the offset was dependent on the pixel location of the
point source.  A position dependent offset was derived by fitting a
3rd order polynomial to the offset as a function of row and column
number.

We observed in the CS ($2 \rightarrow 1$) line during 12 nights in
January of 1993 with the SEST 15~m telescope.  We used the 3 mm schottky
receiver coupled to a 2000 channel acousto--optic spectrometer with a
80 kHz resolution.  The image sideband response was 26 db below that
of the signal.  The object was observed using position switching, with
an integration time of 60 seconds on source followed by 60 seconds on
the reference position.  This was repeated 10 times per position,
giving a total on source integration time of 10 minutes per position
on source.  We mapped the cloud using a $40''$ grid in RA and Dec
(1950). The beamsize was 50\arcsec.

The calibration was made using the chopper wheel method
\citep{2004tra..book.....R}.  The Orion SiO maser was used for pointing.  RMS
pointing uncertainties were less than 8\arcsec.  The total CS($2
\rightarrow 1$) flux of the Orion SiO maser was repeatable to within
10\%, except on the last night.  We note that only 21 points were
observed on the last night: these points were typically
non-detections.  The average system noise (after correcting for the
effect of the atmosphere) was 450 K, the maximum was 590 K. The data
were reduced using the Grenoble CLASS software using standard
procedures.  A beam efficiency of 0.75 was used to convert the antenna
temperatures into main beam temperatures \citep{1994A&A...284...17H};
all our results are on a main--beam temperature scale.

\section{Results}

\subsection{Identification and Classification of Young Stellar Objects}

Young stellar objects (YSOs) with infrared excesses due to dusty
envelopes and disks were identified and classified in G216 using
methodologies developed in a series of publications \citep[for a
  detailed discussion see][]{2007prpl.conf..361A, 2007ApJ...669..493W,
  2008ApJ...674..336G}.  Since these methologies were developed for
star forming regions in the nearest kiloparsec, they were modified to
take into account the greater distance to G216. In addition, a new
method for identifying protostars was also implemented
\citep[also see][]{megeath...in..prep}.  The overall approach can be decomposed
into three steps: the identification of candidate YSOs with
infrared-excesses, the rejection of extragalactic sources
contaminating the sample of candidate YSOs, and finally, the
classification of the remaining YSOs as either protostars or pre-main
sequence stars with disks.  These steps were implemented using a suite
of mid-IR color and magnitude criteria.  In the following overview of
our methodology, we do not entirely segregate these three steps since
the details depend on the specific colors and magnitudes employed.

For the purposes of this paper, we divide our sample of YSOs with
infrared excesses into two gross evolutionary categories: protostars
and young stars with disks.  We effectively classify objects by the
slope of the spectral energy distribution (SEDs). Flat or rising SEDs
with increasing wavelength (i.e. $\alpha \ge -0.3$ where $\alpha =
dlog \lambda F_{\lambda}/dlog\lambda$) are indicative of an infalling
envelope; these include Class 0, Class I and flat spectrum sources.
Decreasing SEDs with increasing wavelengths ($\alpha < -0.3$ ) are
indicative of stars surrounded by circumstellar disks. This approach
is derived from the pre-Spitzer classification scheme of
\citet{1994ApJ...434..614G}.

The starting point of our analysis was the catalog of all point
sources detected by 2MASS, Flamingos, IRAC and/or MIPS; this was
compiled from the point source photometry described in Sec.~2.  The
first steps were to eliminate point sources which were only partially
covered in our {\it Spitzer} observations and to reject photometric
magnitudes with high uncertainties. In this paper, we only consider
the 0.16~sq.~degree region covered by all four IRAC bands as well as
the MIPS 24 and 70~$\mu$m bands; the Flamingos data cover a somewhat
smaller region.  In the following analysis, uncertainties of $\le
0.1$~mag were required for the [3.6] and [4.5] photometry, $\le
0.15$~mag for the [5.8] and [8] photometry, $\le 0.25$~mag for the
[24] photometry and $\le 0.15$~mag for the Flamingos $JHK$ photometry.
Sources were kept in the inital point source catalog as long as they
contained at least one photometric magnitude

\begin{figure}
\epsscale{1.}
\plotone{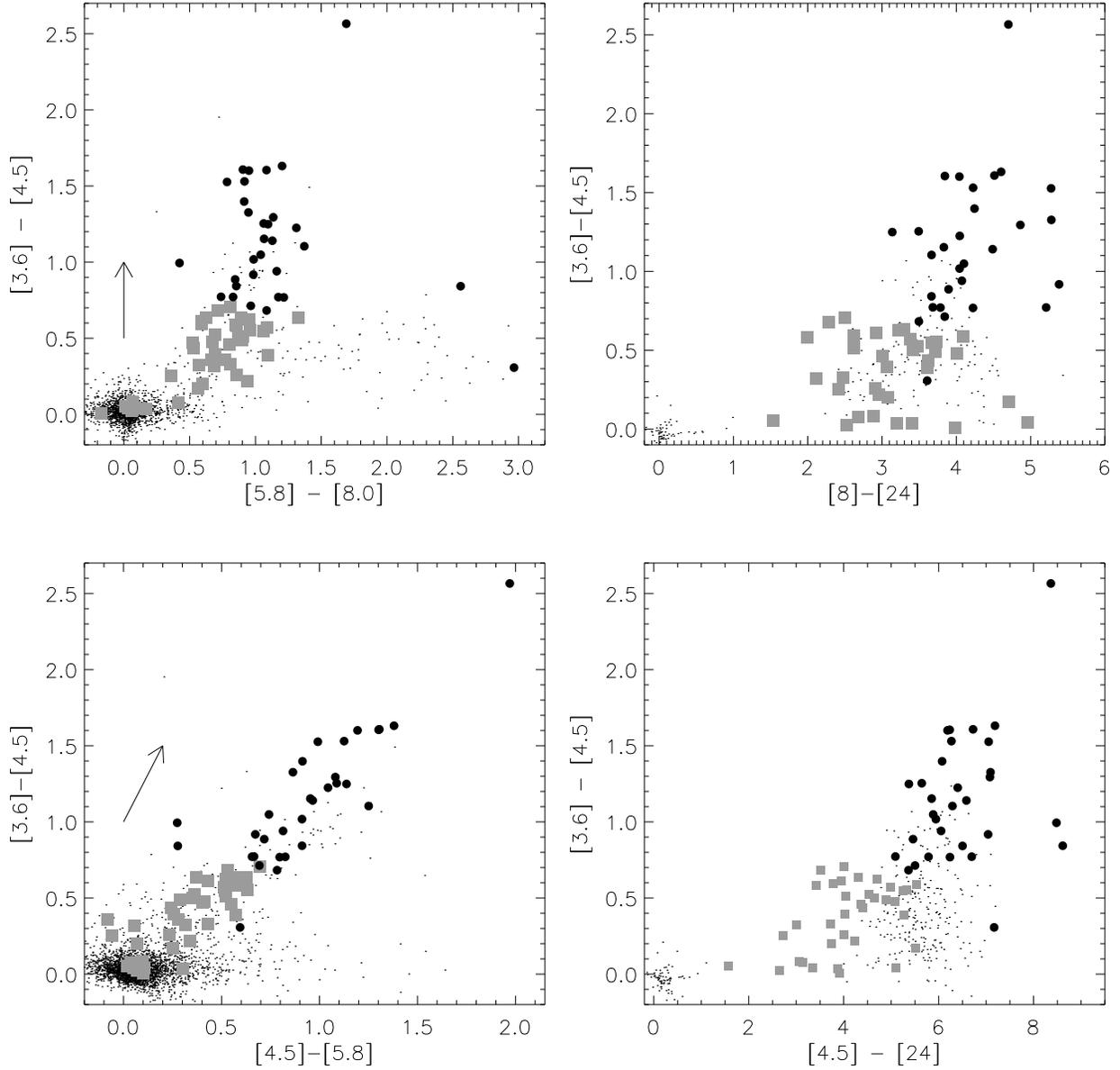}
\caption{Color-Color Diagrams.  All sources with photometry satisfying
  the uncertainty requirements are plotted as dark dots.  Those
  classified as protostars are circles, and those as stars with disks
  are squares.  The vectors represent 5 magnitudes of
  extinction in the $K$-band using the reddening law of
  \citet{2007ApJ...663.1069F}.}
\label{fig:cc}
\end{figure}

Using models of protostars and young stars with disks,
\citet{2004ApJS..154..363A} and \citet{2004ApJS..154..367M} found the
IRAC [3.6]-[4.5] vs. [5.8]-[8] diagram a powerful means for
identifying infrared excess sources (Fig.~\ref{fig:cc}); these sources
include both protostars and young stars with disks.  We select
IR-excess sources which satisfy the following criteria:

\begin{equation}
[3.6]-[4.5] \ge 0.2 + \sigma_{[3.6]-[4.5]}, ~~ [5.8]-[8.0] \ge 0.35 + \sigma_{[5.8]-[8]} 
\end{equation}

\noindent
Many sources can be detected in the IRAC 3.6~$\mu$m and 4.5~$\mu$m
bands but are not detected in the longer wavelength IRAC and MIPS
bands because of the lower sensitivity to photospheres in these bands.
\citet{2004ApJS..154..374G} and \citet{2007ApJ...669..493W} used
combined IRAC and near-IR color-color diagrams to identify young
stellar objects that lack detections longward of 4.5~$\mu$m.  However,
given our exceptionally sensitive 5.8 and 8~$\mu$m data (compared to
other surveys of star forming regions - this results from our
relatively long integrations and the lack of a bright nebulous
background), the comparatively low sensitivity of the 2MASS data for
this distant region, and the $\sim 0.1$~mag. scatter in the deeper
Flamingos photometry, we find the combined IRAC/near-IR diagrams do
not show reliable new YSOs.  We do find that the addition of the
24~$\mu$m data does result in the identification of new YSOs.  In
particular, young stars with disks may not show strong excesses in the
IRAC bands due to holes in their inner disks; these are the transition
disks sources \citep{2004ApJS..154..379M}.  Among the sources with
detections in the MIPS 24~$\mu$m band, we search for young stars with
transition disks by using the color criteria \citep[Fig.~\ref{fig:cc},
  also see][]{2007ApJ...669..493W}:

\begin{equation}
[8]-[24] > 1
\end{equation} 

\noindent

\begin{figure}
\epsscale{1.}
\plotone{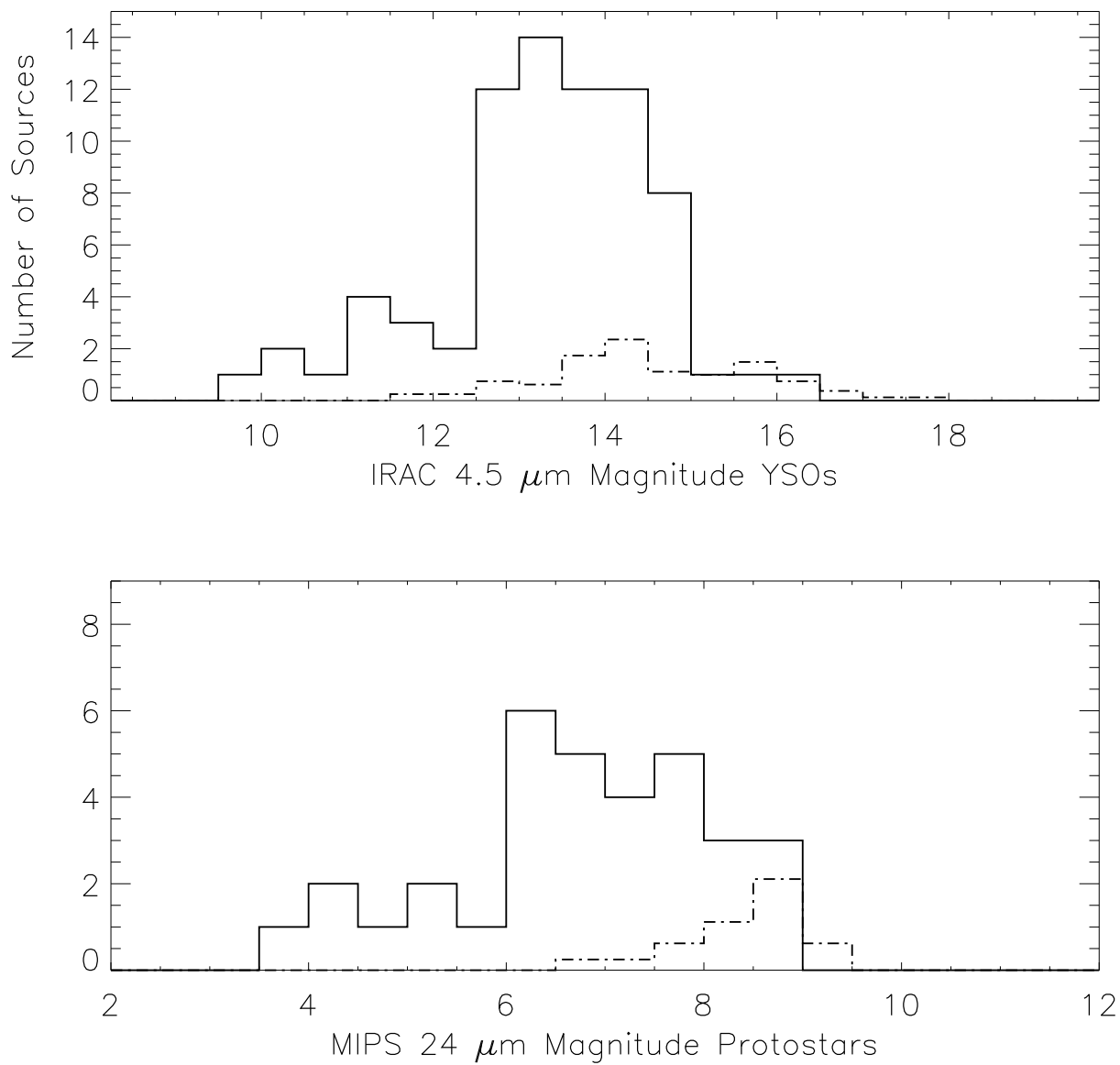}
\caption{{\bf Top:} Histogram of expected AGN contamination as a
  function of 4.5~$\mu$m magnitude.  Solid line: distribution of
  4.5~$\mu$m magnitudes for our sample of YSOs in G216.  Dot-dashed
  line: the estimated distribution for extragalactic sources
  misidentified as YSOs based on the SWIRE data. {\bf Bottom:}
  Histogram of expected galaxy contamination to our protostar sample.
  Solid line: distribution of 24~$\mu$m magnitudes for our samples of
  protostars.  Dot-dashed line: the estimated distribution of
  misidentified extragalactic sources, again based on the SWIRE data.}
\label{fig:hist}
\end{figure}

The sample of infrared excess objects selected by the above criteria
is contaminated by galaxies with strong PAH emission and AGN
\citep{2005ApJ...631..163S}.  PAH dominated galaxies exhibit colors
distinct from those of YSOs; we eliminate such galaxies using the
empirically derived color criteria given in the appendix of
\citet{2008ApJ...674..336G}.  In contrast, AGN show colors similar to
those of YSOs and must be distinguished by their fainter magnitudes.
We determined the optimal threshold magnitude by examining the density
of AGN in survey data from the SWIRE legacy program; the SWIRE fields
were chosen to study extragalactic populations with a minimum of
contamination from galactic objects \citep{2003PASP..115..897L}. We
used the point source catalog the from Elias-N2 field obtained from
the NASA/IPAC Infrared Science Archive. In Fig.~\ref{fig:hist} we show
the number of AGN with colors satisfying our YSO color criteria
(Equations 1 \& 2) as a function of their 4.5~$\mu$m magnitude.  The
number is corrected for the smaller size of the G216 field. Assuming
that the density of AGN is the same in the SWIRE and G216 fields, the
histograms show that the number of expected AGN exceeds the number of
YSOs for $[4.5] \ge 15$.  However, instead of adopting a constant
magnitude threshold, we use the color dependent magnitude threshold of
\citet{2008ApJ...674..336G}.  We retain the color dependence of the
Gutermuth et al. criteria, but we lower the magnitude threshold by
0.5~mag to account for the greater distance to G216.  Consequently,
for a source to be classified as an AGN, it must satisfy all of the
following criteria:

\begin{eqnarray*}
& [4.5] -[8.0] > 0.5 \\
& [4.5] > 14.+([4.5]-[8.0]-2.3)/0.4\\
& [4.5] > 14
\end{eqnarray*}

\noindent
In addition, a source would have to satisfy one of the following three
criteria before it is classified as an AGN:

\begin{eqnarray*}
& [4.5] > 14.5+([4.5]-[8.0]-0.5)  \\
& [4.5] > 15 \\
& [4.5] > 15- ([4.5]-[8.0] -1.2)/0.3
\end{eqnarray*}

\noindent
Sources with colors consistent with Equations~1 and 2 but satisfing
the above criteria are rejected from the YSO sample.  The division
between sources selected as YSOs and AGN is shown in the [4.5]-[8]
vs. [4.5] diagram displayed in Fig.~\ref{fig:cm}.  This figure shows
that the adopted threshold is typically brighter than 15~mag at 4.5~$\mu$m,
and consequently, will reject the vast majority of contaminating AGN
(Fig.~\ref{fig:hist}).  Finally, to minimize galactic contamination in
the sample of transition disks (which do not have the colors of AGN),
we deleted one transition disk candidate with $[4.5] > 15$~mag.

The above criteria result in a sample of YSOs containing a mixture of
protostars and young stars with disks.  To distinguish between
protostars and young stars with disks in that sample, and to further
increase the sample of protostars, we apply criteria utilizing only
the three most sensitive bands for detecting protostellar sources: the
3.6~$\mu$m, 4.5~$\mu$m and 24~$\mu$m bands.  Protostars may not be
detected in the 5.8~$\mu$m and 8~$\mu$m bands because of the lower
sensitivity in these bands, the bright nebulosity from the strong
hydrocarbon features between 5 and 9~$\mu$m, and the flattening and
dip in the protostellar SED apparent in the 5-10~$\mu$m spectral
regime.  Furthermore, protostars may not be detected in the near-IR
bands due to their high extinction.

We identify protostars by adopting the criteria that objects
exhibiting SEDs with spectral indices $\alpha \ge -0.3$ are protostars
\citep{1994ApJ...434..614G,megeath...in..prep}.  We implement this
criteria by determining the photometric colors of a power-law SED with
$\alpha \ge -0.3$ using the IRAC and MIPS spectral response curves
posted on the {\it Spitzer} Science Center website. We find that
protostars exhibit the following colors (Fig.~\ref{fig:cc}):

\begin{equation}
[3.6]-[4.5] \ge 0.652,~~[4.5]-[24] \ge 4.761
\end{equation}

\begin{figure}
\epsscale{1.}
\plotone{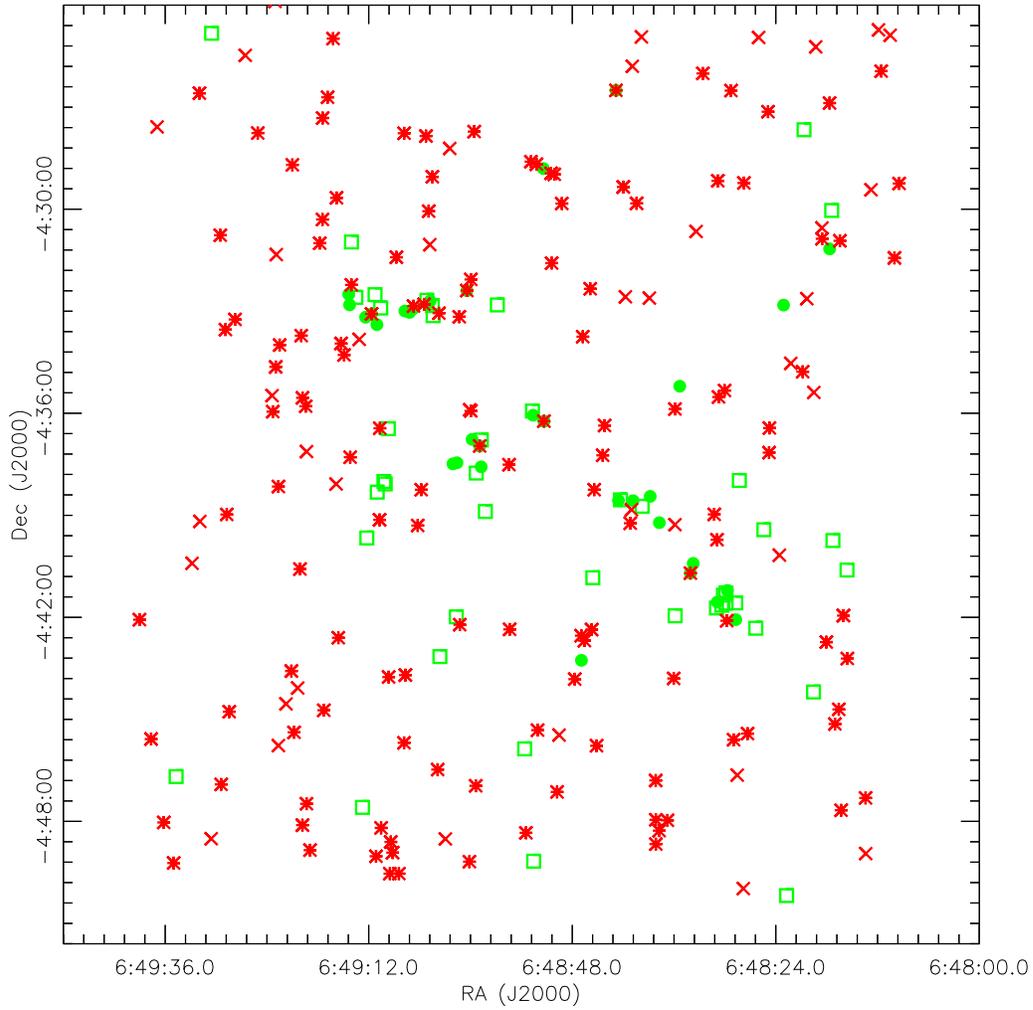}
\caption{Distribution of candidate YSOs, AGN, and star-forming galaxies.
Green filled circles are protostars, green squares are young stars with 
disks, red asterisks are AGN and red Xs are galaxies identified by their
their faint 24~$\mu$m magnitudes.}
\label{fig:cont}
\end{figure}

The sample of protostars selected by these color criteria also suffers
from contamination by galaxies. Specifically, the [4.5]-[24] vs. [24]
diagram (Fig.~\ref{fig:cm}) shows a distinct clump of faint sources
that satisfy our color criteria; these sources are scattered evenly
across the sky and are probably extragalactic (Fig.~\ref{fig:cont}).
In Fig.~\ref{fig:hist}, we plot the the number of galaxies in the
SWIRE field which satisfy the protostar color criteria (Eqn.~3) as a
function of 24~$\mu$m magnitude; once again the number has been scaled
to account for the smaller size of the G216 field.  We compare this to
the number of protostars candidates identified in the G216 field. We
find that for $[24] < 9$, the number of protostar candidates exceeds
the number of expected galaxies.  To minimize contamination from
galaxies, we adopt the following empirical criteria for YSO:

\begin{equation}
[24] \le 9.15
\end {equation}

\noindent
We find one faint protostar candidate ([24] = 9.1~mag.) clustered with
the brighter protostars; we have set the threshold for YSOs to
9.15~mag so that this source is included in our YSO list.  The
application Equations~3 \& 4 results in a sample
of protostars with the following three pedigrees:

\begin{enumerate} 
\item{} objects {\it identified} as YSOs using the previous critiera but not previously
 {\it classified} as protostars

\item{} resurrected YSOs which were previousy rejected as AGN but are
  now considered protostars (since deeply embedded protostars can have
  weak 4.5~$\mu$m emission, the rejection of galaxies from the
  protostar sample is better done at 24~$\mu$m).

\item{} newly identified protostars which were not previously identified as infrared excess
sources because of a lack of [5.8] and/or [8] photometry.  
\end{enumerate} 

\noindent

Although limited by a lower angular resolution and sensitivity than
the other {\it Spitzer} bands, the 70~$\mu$m data adds an additional
means to identify cold protostars.  One additional source is
identified as a protostar on the basis of the 70~$\mu$m data. This
source was not initially identified as a protostar since its
[3.6]-[4.5] color is similar to a young star with disk. Because of its
faintness and its color, it was initially classified as an AGN.
However, given its brightness in the 70~$\mu$m band and its extremely
red color, we include this in our protostar sample (source 14 in
Table~1). The remaining sources with 70~$\mu$m band detections were
not classified as YSOs; these sources are fainter in the 70~$\mu$m
band than the previously identified YSOs and have a 70~$\mu$m
magnitude $> 3$.  In comparison, only one previously identified YSO
has a 70~$\mu$m magnitude $>3$; this was identified as a protostar
through its 3.6~$\mu$m, 4.5~$\mu$m and 24~$\mu$m photometry (source 12
in Table~1).  To further test whether the unclassified sources with
[70] $> 3$~mag are background galaxies or YSOs, we examine their
spatial distribution.  We find that these faint sources are
distributed uniformly in the 70~$\mu$m mosaic; this is in distinct
contrast to the protostellar sources with 70~$\mu$m magnitudes $\le
3$ which are concentrated in the center of the field.  We
conclude that the unclassified sources with [70] $> 3$ mag are primarily
galaxies and do not include them in our YSO sample.

In total, 33 protostars were identified (Table~1).  An alternative
approach to classify protostars using the [4.5]-[5.8] colors of YSOs
was proposed by \citet{2008ApJ...674..336G}.  With only three
exceptions, the colors of our protostars are consistent with this
other criteria (Fig.~\ref{fig:cc}).  The remaining YSOs with infrared
excesses which were not classified as protostars were classifed as
young stars with disks. A total of 41 stars with disks were identified
(Table~1). The distribution of protostars and young stars with disks
are shown in Fig.~\ref{fig:dist}.

The criteria described above are designed to minimize the
contamination of the YSO sample by galaxies. The success of these
criteria are apparent in Fig.~\ref{fig:cont}, where we show the
distribution of the YSOs and the identified contaminating galaxies.
While the galaxies are uniformly distributed across the field, the
YSOs do not exhibit the same uniform distribution.  Nevertheless, we
expect a residual of galaxies in the YSO sample.  To determine the
level of residual extragalactic contamination, we applied the full
suite of criteria described above to the SWIRE catalog for the Elias-N2
field.  After the rejection of likely galaxies, we estimate the
remaining contribution of extragalactic contamination to our sample is
approximately 6.2 protostars and 7.5 stars with disks.  After the
subtraction of this background, the corrected numbers of bona-fide
protostars and young stars with disks are 27 and 33.5, respectively.
We note that most of the protostars contamination occurs at the
fainter magnitudes, with half of the extragalactic sources
misidentified as protostars having [24] magnitudes between 8.65 and
9.15 (Fig.~\ref{fig:hist}).

\subsection{Distribution of Young Stellar Objects and Dense Cores}

\begin{figure}
\vskip -0.5 in
\epsscale{0.7}
\plotone{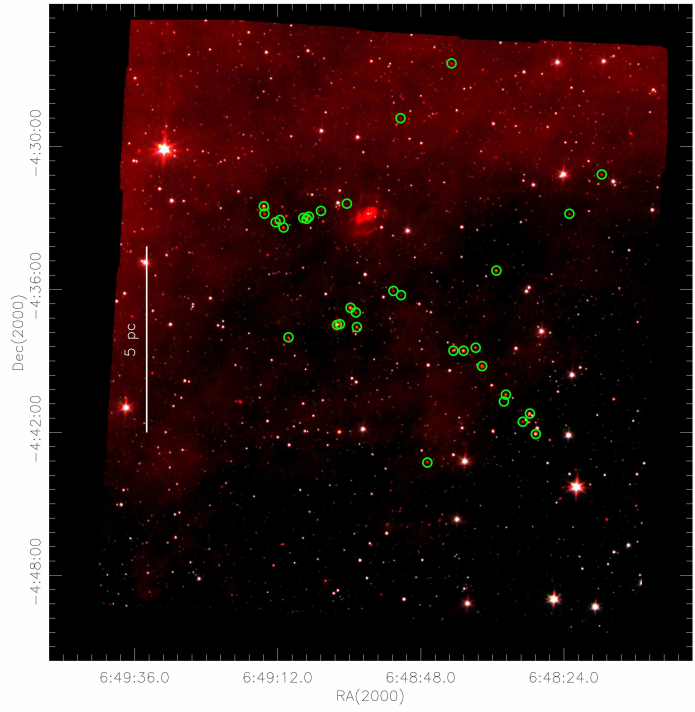}
\plotone{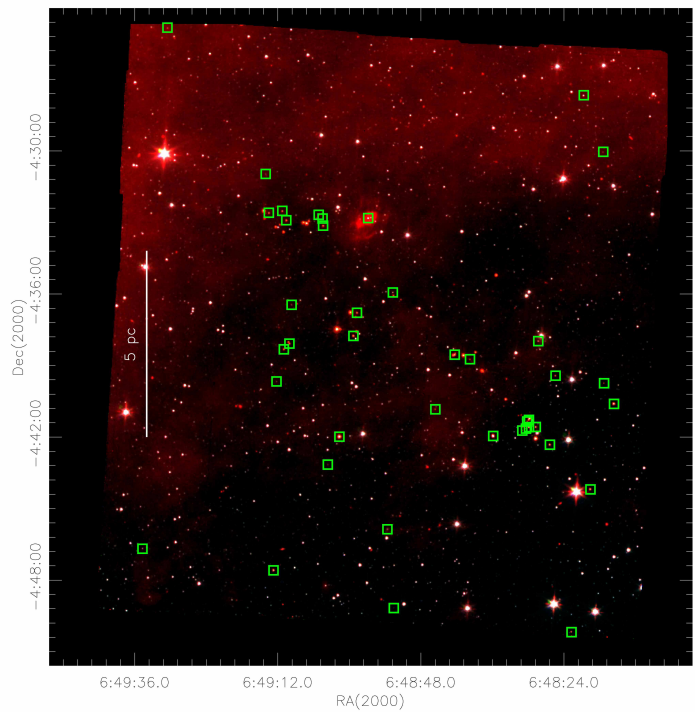}
\caption{{\bf Top:} the distribution of protostars. {\bf Bottom:} the
distribution of young stars with disks.  The background image is the
combined IRAC 3.6 (blue), 4.5 (green) and 8~$\mu$m (red) image.  }
\label{fig:dist}
\end{figure}

\begin{figure}
\epsscale{1.}
\plotone{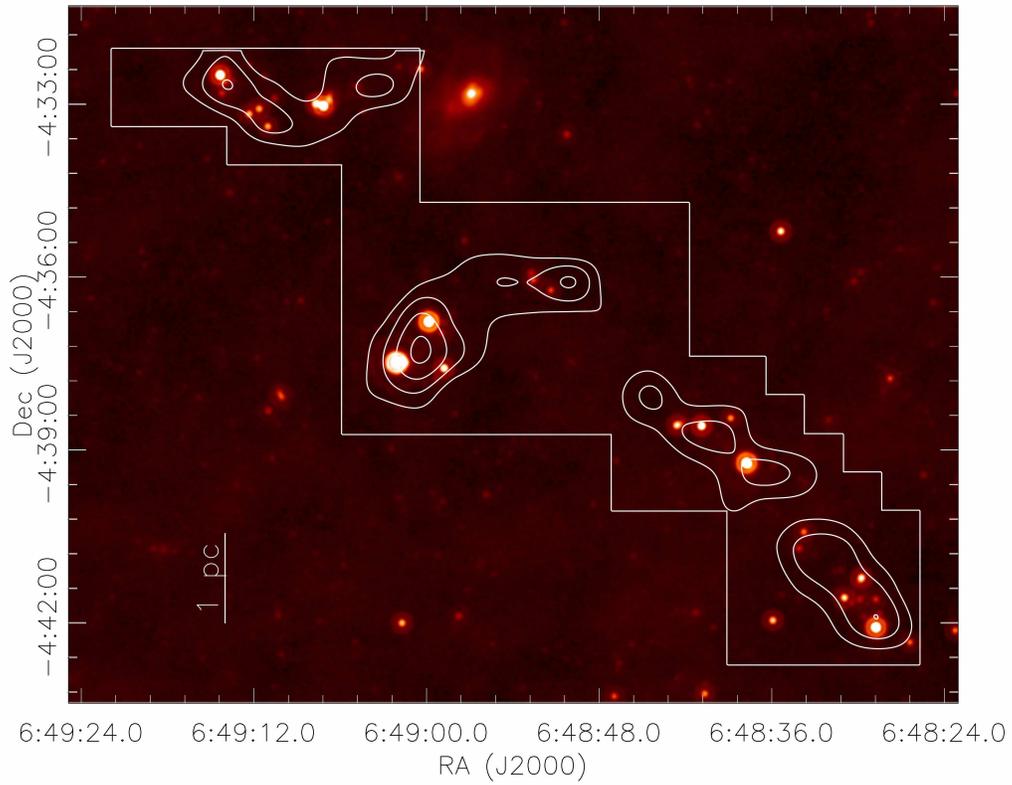}
\caption{MIPS 24 micron image of the central region containing the
young stellar objects. The contours show the velocity integrated CS
($2 \rightarrow 1$) emission; the contour levels are 0.75, 1, 1.25 and
1.5~K~km~s$^{-1}$. The extent of the CS map is also shown (delineated
by the straight lines).  The 1 pc scalebar is shown for the adopted 
distance of 2.2 kpc.}
\label{fig:cs}
\end{figure}

In Fig.~\ref{fig:dist}, we show the distribution of young stellar
objects overlaid on the IRAC images of the region.  The MIPS 24~$\mu$m
image is displayed in Fig.~\ref{fig:cs}; overlaid on the MIPS data are
contours of velocity integrated CS ($2 \rightarrow 1$) emission. The
extent of the CS map, which is smaller than the 24~$\mu$m mosaic, is
also displayed. The 33 protostars and 41 stars with disks are
concentrated in a region which extends diagonally through the map.
The dense gas traced by the CS ($2 \rightarrow 1$) emission forms a
clumpy ridge extending 10~pc diagonally across the field. Most
of the protostars are coincident with detectable CS emission; the six
exceptions are outside the region mapped with the SEST.  Most of the
stars with disks are also concentrated near the dense gas ridge; the
remainder are found to the south of the ridge.  The region north of
the ridge is almost devoid of YSOs.  The number of protostars outside
the ridge is equal to the number of expected contaminating galaxies;
this suggests that the protostars are confined to the ridge with the
sources outside the ridge being likely misidentified galaxies.

\begin{figure}
\epsscale{1.}
\plotone{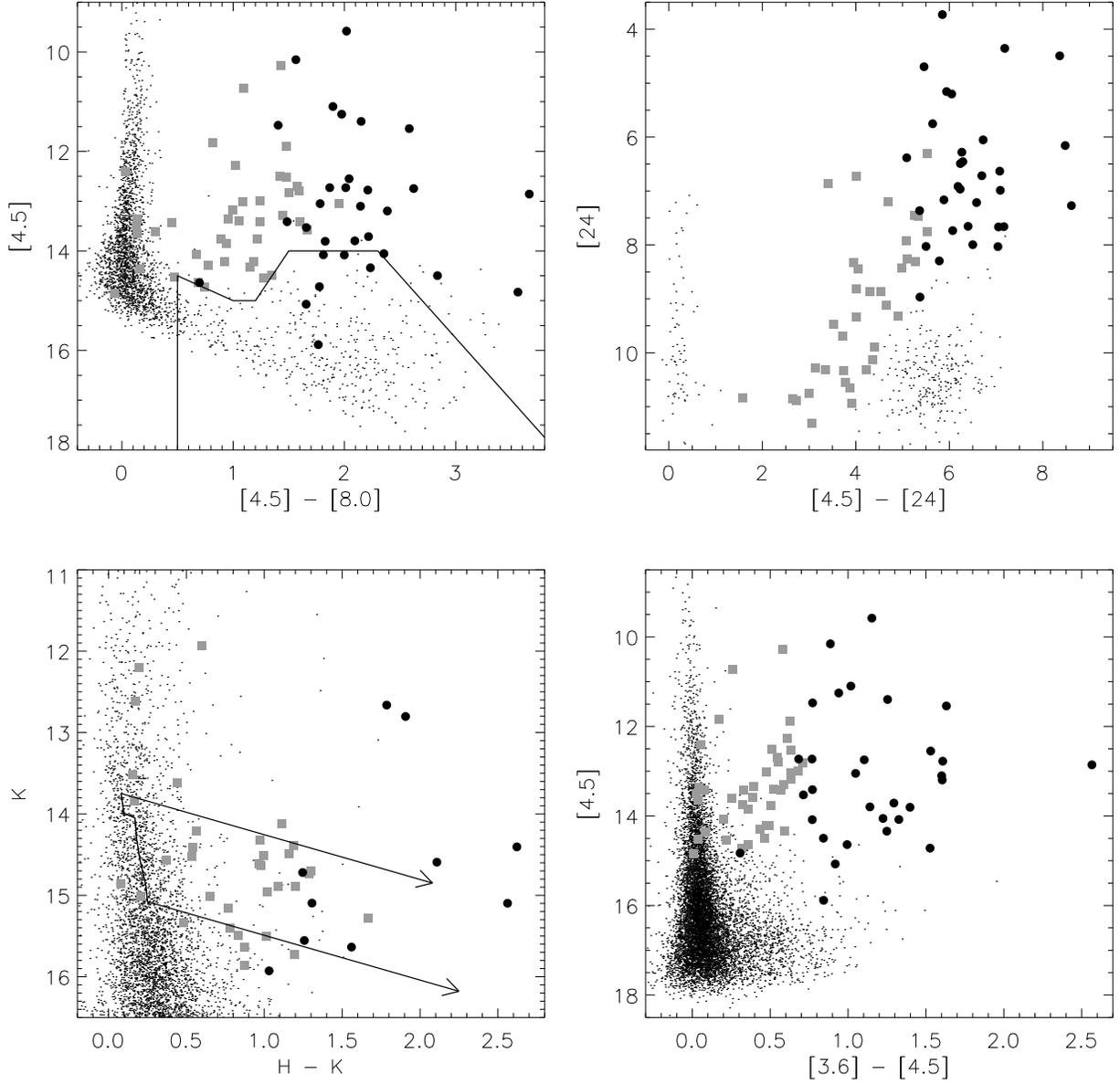}
\caption{Color-Magnitude Diagrams.  Symbols are same as Fig.~2.  The
line drawn on the [4.5]-[8.0] vs. [4.5] plot represents the criteria
used for eliminating AGNs from the data set. Note that only stars with
disks were required to satisfy this criteria. A large concentration of
faint, red extragalactic sources is also apparent in the
[4.5]-[24] vs. [24] plot (centered on [24] $\sim 10.5$ and [4.5]-[24] $\sim
5.5$).  The $H$ vs. $H-K$ diagram shows the 1 Myr isochrone from
\citet{1998A&A...337..403B}; the upper and lower arrows are reddening
vectors corresponding to 10 A$_V$ at masses of 1.2~M$_{\odot}$ and
0.25~M$_{\odot}$, respectively.  The [3.6]-[4.5] vs [4.5] diagram show
the photometry of the YSOs relative to all sources detected in the [3.6]
and [4.5] bands.
}
\label{fig:cm}
\end{figure}

The H-K vs. K plot in Fig.~\ref{fig:cm} shows the
\citet{1998A&A...337..403B} pre-main sequence tracks for 1 Myr YSOs
plotted over the G216 sources.  Adopting an age of 1~Myr, most of the
pre-mains sequence stars would have masses of 0.25-1.2~M$_{\odot}$
(Fig.~\ref{fig:cm}), with the faintest identified stars having masses
close to the Hydrogen burning limit.  A compact red nebula is apparent
in Fig.~\ref{fig:dist}; the nebula exhibits bright 8~$\mu$m emission
indicative of UV heated hydrocarbons.  The UV source is likely to be a
young star with disk found within the nebula (source 56 in Table~1).
Interestingly, the infrared excess from this star is weak except at
24~$\mu$m.  The photometry may be affected by the nebular emission;
however, visual inspection of the PSF subtracted 24~$\mu$m mosaic
shows that the photometry is not significantly contaminated by the
nebula.  The star is one of the more luminous pre-main sequence stars
in this region of the cloud.  Adopting an age of 1~Myr, the bright
source toward the nebula lies well above the 1.2 solar mass upper
limit for the tracks in Fig.~\ref{fig:cm}.  Two others star/disks
(sources 38 and 59 in Table 1) have similar near-IR luminosities.  A
comparison of the photometry of these sources with Herbig Ae/Be stars
from \citet{1992ApJ...397..613H} shows that these sources are
consistent with A type stars, with K-band magnitudes and H-K
colors similar to the A2 star HD 245185.  Although spectroscopy is needed to
confirm spectral types, these data suggest that the most massive young
stars in the region are early A~type stars.  This would explain why
the central core of G216 is not prominent in IRAS maps of the far-IR
emission.

\section{Discussion}

Although these observations clearly demonstrate that star formation is
ongoing in the central core of G216, the 74 YSOs are both small in
number and distributed over a region of 230 pc$^2$.  The high
proportion of protostars indicates that this is a young population of
stars, and is not the result of the past episode of star formation
proposed by \citet{1994ApJ...432..167L}. However, such an older
population may not exhibit the bright infrared excesses used to
identify the YSOs in this study, and we cannot rule out its existence.

How does this compare to other molecular clouds? We compare G216 to
two of the best studied molecular clouds near the Sun: the Taurus dark
cloud complex, the prototype cloud for distributed star formation, and
the Orion~A giant molecular cloud, the prototype cloud for clustered
star formation.  Since a comprehensive {\it Spitzer} catalog for
Taurus has not been published, we cannot estimate the number of YSOs
with IR-excesses that would be selected in a 2200~pc distant Taurus
cloud.  Instead, we take a recent compilation of all known YSOs in
Taurus (K. Luhman, P. Com); this contains a total of 296 sources. The
entire Taurus cloud is extended over a region 1408~sq~pc; the YSOs
fill only a fraction of this region \citep{1987ApJS...63..645U}. For
Orion~A, we use the {\it Spitzer} Orion survey and apply similar
criteria as applied to G216 for identifying YSOs with IR-excesses
\citep{megeath...in..prep}.  A total of 1935 YSOs are identified in
the {\it Spitzer} survey, which covers 261 sq pc
\citep{megeath...in..prep}.

\begin{figure}
\epsscale{1.}
\plotone{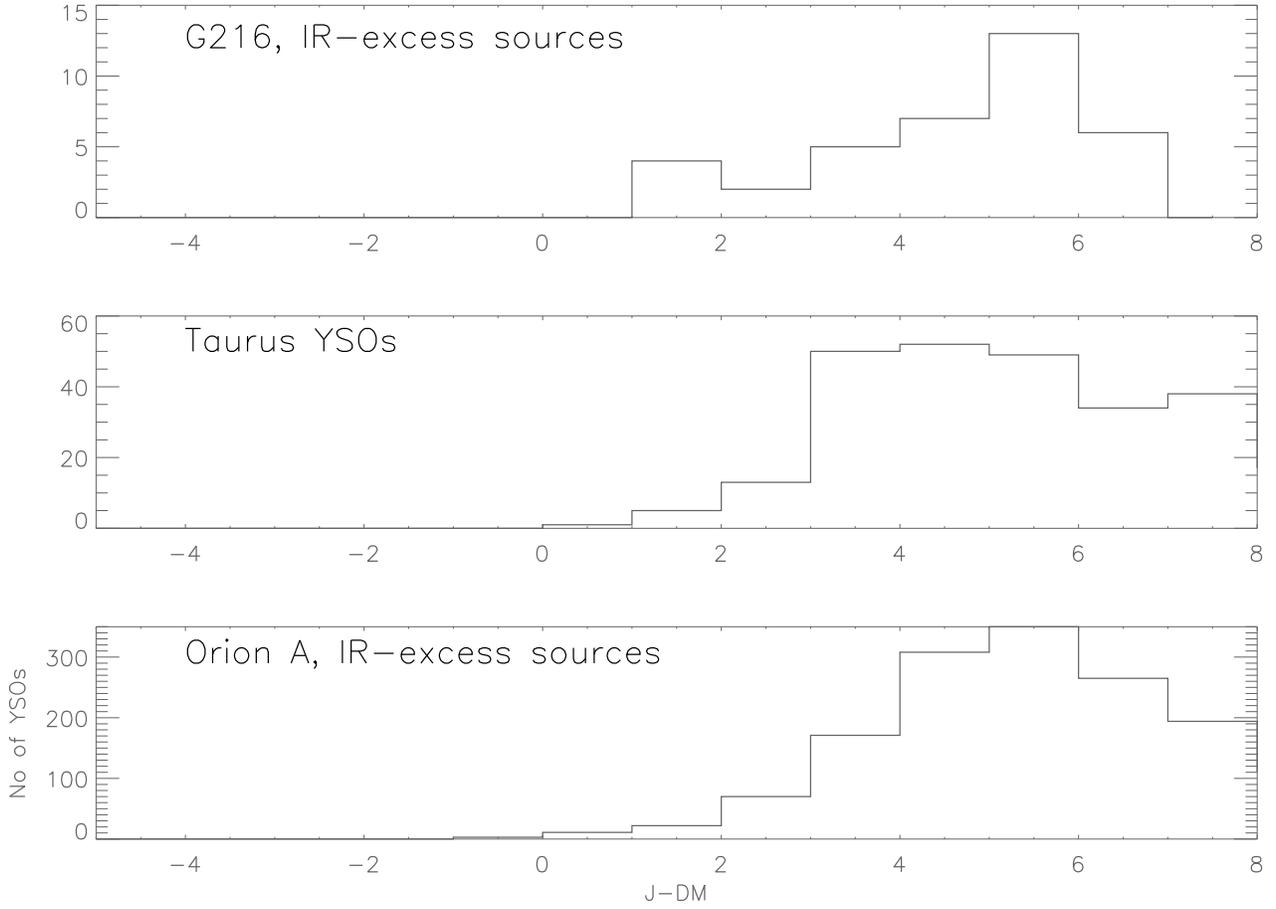}
\caption{Histograms of the $J$-band magnitudes for the Taurus, Orion~A
and G216 samples.  The $J$-band magnitudes have been corrected for the
distance modulus (DM), adopting distances of 142~pc, 414~pc and 2200~pc for
Taurus, Orion and G216 respectively.  The faintest source in the G216
sample has a $J-DM$ = 6.15 (Table 1).}
\label{fig:comp_cc}
\end{figure}

We now compare the densities of YSOs in the G216, Orion and Taurus
samples.  To compare these three samples, we must account for the
larger distance of G216. This can be done by invoking a cutoff in the
$J$-band magnitude equivalent to the faintest $J$-band magnitude of
the G216 YSO sample.  From Fig.~\ref{fig:comp_cc}, this value is $J-DM
\le 6.15$, where $DM$ is the distance modulus.  Sources undetected in
the $J$-band, typically prototstars, were not eliminated.  This is not
a rigorous determination of the completeness in the G216 data;
however, this crude approach is sufficient for our comparison of
stellar densities. Using the $J$ magnitude cutoff, we estimate that we
would detect 1372 sources in the Orion~A cloud if it were placed at
2200 pc.  In Taurus, we would detect 183 sources at 2200 pc.

\begin{figure}
\epsscale{1.}
\plotone{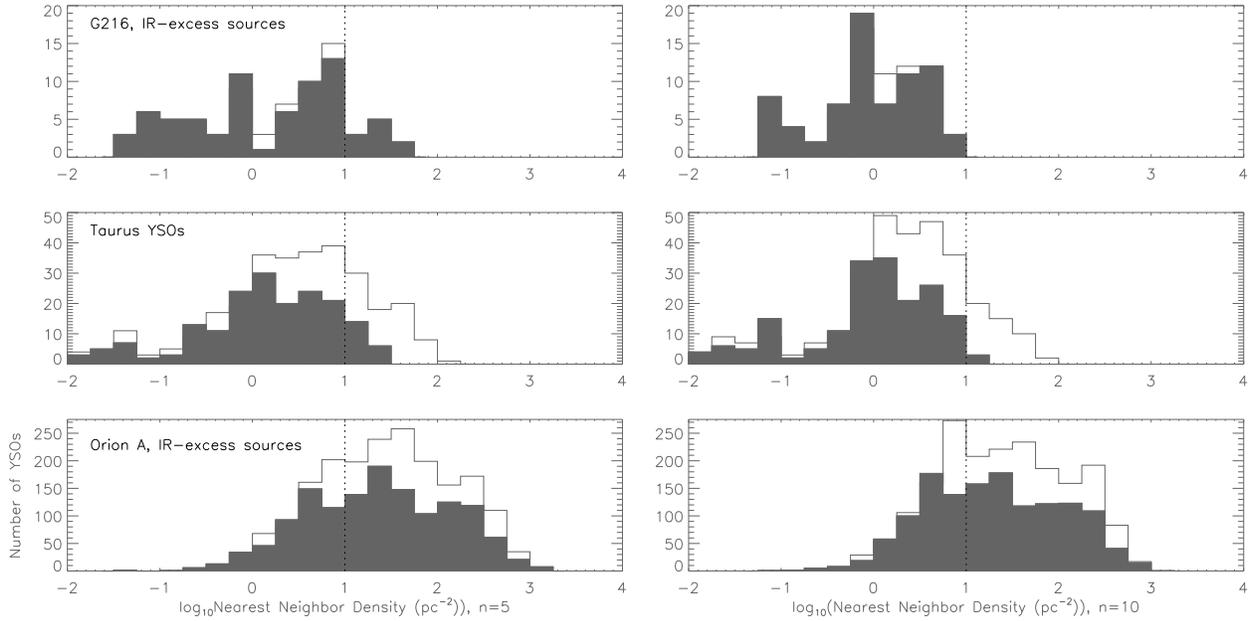}
\caption{Histograms of YSO nearest neighbor densities for G216, Taurus
  and Orion~A.  The densities are calculated for each YSO using the
  equation density = $n/(\pi r_n^2)$ where $r_n$ is the distance to
  the nth nearest YSO and $n$ is set to 5 (left) and 10 (right).  The
  upper line in each plot outlines the histogram using all the YSOs in
  each of the three clouds.  The shaded histograms use only YSOs that
  satisfy the condition $J-DM \le 6.15$~mag. or are not detected in
  the $J$-band. The dashed line marks the density of 10~pc$^{-2}$;
  \citet{megeath...in..prep} define groups and clusters as regions
  with 10 or more YSOs that have a surface density in excess of this
  threshold.  Consequently, G216 does not contain a group or cluster
  by this definition.}
\label{fig:comp_nn}
\end{figure}

In~Fig.~\ref{fig:comp_nn}, we show the distribution of local YSO
surface densities measured around each YSO for G216, Taurus and
Orion~A. The surface density was calculated using the nearest neighbor
method, in which $density = n/(\pi r_n^2)$ where $r_n$ is the distance
to the nth nearest YSO
\citep{2005ApJ...632..397G,1985ApJ...298...80C}.  This was done for
$n=5$ and $n=10$ using the total sample of YSOs in each cloud and
using the sample of YSOs that satified the $J-DM$ criteria.  The
resulting surface densities in G216 are much lower than the Orion A
cloud, but are similar to those found in the Taurus dark cloud
complex.

The density and configuration of YSOs in G216 are similar to those in
the Taurus cloud; with small, low density groups of stars distributed
among filamentary clouds.  However, the cloud mass of G216 is ten
times that of the Taurus cloud and is comparable to that of the
Orion~A cloud \citep{2005A&A...430..523W}.  Interestingly, the
properties of the turbulence in G216 seems to be similar to more
active star forming GMCs \citep{2006ApJ...643..956H}.  It is, however,
unlikely that G216 will evolve into an Orion cloud given the estimated
star formation rate.  Assuming a 0.43~Myr year lifetime for protostars
\citep{2007A&A...468.1009H}, we estimate a star formation rate of 60
stars per Myr for G216. Thus, at this rate, it would take 17~Myr for
G216 to form 1000 stars.  This is longer than the measured lifetimes
of molecular clouds \citep{2001ApJ...562..852H}.  G216 could produce a
Taurus-like region in 3 million years. The CS emission in G216
exhibits lower brightness temperatures and slightly lower linewidths
than those typically exhibited by star forming dense cores in GMCs
\citep{1994ApJ...432..167L,1998ApJ...494..657W}.
\citet{1994ApJ...432..167L} suggested that the G216 cores are similar to
those in Taurus; however, a direct comparison of the dense gas is complicated
by the low spatial resolution (0.6~pc) of the G216 CS ($2 \rightarrow
1$) map.  We defer this analysis to a future paper using existing
multi-transition CS data.

\section{Summary}

We present the first detection of ongoing star formation in the main
body of G216, otherwise known as Maddalena's cloud or the
Maddalena-Thaddeus cloud.  We identify 33 protostars (6.2 of which are
expected to be contamination) and 41 young stars with disks (7.5 of
which are expected to be contamination). This demonstrates that G216
is not quiescent, but forming young stars at an approximate rate of 60
stars per million years.  The most luminous pre-main sequence stars
have near-IR luminosities similar to Herbig Ae stars, consistent with
the lack of massive star formation in the cloud inferred from IRAS
far-IR maps. The star formation is concentrated in a 10~pc long
molecular ridge detected in CS ($2 \rightarrow 1$) observations. The
density of star formation is much lower than that found in more active
GMCs such as Orion GMCs, and is similar that in the Taurus dark clouds.  This may
result from the presence of less dense gas and smaller dense cores
compared to active star forming regions such as Orion.  Future work
should concentrate on detailed comparisons of the molecular cores of
G216 with those in Taurus, Orion and other molecular clouds, with the
goal of understanding how the properties of the gas determines the
rate and density of star formation.

\acknowledgments

This work is based on observations made with the {\it Spitzer} Space
Telescope, which is operated by the Jet Propulsion Laboratory,
California Institute of Technology, under NASA contract 1407.  Support
for this work was provided by NASA through contract Number 1285132
issued by JPL/Caltech.  This publication makes use of data products
from the Two Micron All Sky Survey, which is a joint project of the
University of Massachusetts and the Infrared Processing and Analysis
Center/California Institute of Technology, funded by the National
Aeronautics and Space Administration and the National Science
Foundation. This research has made use of the NASA/IPAC Infrared
Science Archive, which is operated by the Jet Propulsion Laboratory,
California Institute of Technology, under contract with the National
Aeronautics and Space Administration.  We thank the referee Bruce 
Wilking for  prompt and very helpful comments.

\clearpage

\def\newblock{}
\def\aap{AA}
\def\aj{AJ}
\def\apj{ApJ}
\def\apjl{ApJL}
\def\apjs{ApJS}
\def\baas{BAAS}
\def\nat{Nature}
\def\mnras{MNRAS}
\def\apss{APSS}
\def\araa{ARAA}
\def\angst{\AA}

\bibliography{ADS}

\end{document}